\documentclass[aps,pra,preprint,amssymb, amsfonts,amsmath,showpacs, nofootinbib,superscriptaddress]{revtex4-1}
\usepackage{graphicx}
\usepackage{amssymb}
\usepackage{bbold}

\begin{document}
  
\title{Relativistic Collapse Model with Quantised Time Variables}
  
\author{Daniel J. Bedingham}
\email{daniel.bedingham@rhul.ac.uk }
\affiliation{Department of Physics, Royal Holloway, University of London, Egham, TW20 0EX, UK}

\author{Philip Pearle}
\email{ppearle@hamilton.edu}
\affiliation{Emeritus, Department of Physics, Hamilton College, Clinton, NY  13323}
\begin{abstract}
 {    
 A relativistic collapse model for distinguishable particles is presented. Position and time, for each particle, are the fundamental operators of the theory. 
 The Schr\"odinger equation is of the CSL form, with a Hermitian Hamiltonian and an anti-Hermitian, white-noise dependent, Hamiltonian. It generates state vector evolution parametrised by an ``evolution parameter." 
 It is shown how this can be interpreted as an evolving state in spacetime with collapses satisfying Born rule probabilities, and how certain choices of collapse generating operators lead to  states of definite mass and definite configuration in spacetime. The model is Poincar\'e covariant and conserves energy in expectation.
 
 }
 
 \end{abstract}

\maketitle

\section{Introduction}\label{}

In standard quantum theory, a measurement invokes the {\it collapse postulate}, an ill-defined  discontinuous  transition of the state vector to a new state vector which reflects the observed measurement outcome. The theory includes no mathematically precise definition of what constitutes a measurement, nor when the transition is to take place. This is widely accepted as a flaw in the theory, and is commonly known as the measurement problem.

Collapse models are attempts to construct a more fundamental state vector dynamics that resolves the measurement problem.  The Schr\"odinger equation evolution is modified.  When describing a measurement situation, the new dynamics causes the state vector to evolve from the superposition of post-measurement  states given by the usual Hamiltonian dynamics to one or another of these states. The modification is expected to be stochastic since measurement outcomes are random, and it is required that the probability of any given outcome is given by the Born rule.

For example, in the SL (Spontaneous Localisation) model of Ghirardi, Rimini and Weber~\cite{GRW}, the modification corresponds to a random sequence of imprecise particle position measurements. There is no suggestion that measurements in the usual sense are taking place; the state changes spontaneously in all physical situations. These abrupt changes of the state vector occur against the background of the usual Schr\"odinger dynamics and it is arranged that the random changes are so rare that they have barely any effect on any single particle. For small systems the dynamics is then well approximated by the Schr\"odinger equation. 

When a quantum measurement is performed, the situation evolves to the creation of an entanglement between the state of a small quantum system and the macroscopic state of a measuring device, such as a physical pointer. When each constituent particle of the macroscopic pointer is subjected to these random collapses, collapse on just one of these many particles rapidly leads to the collapse to one of the pointer states and, by the entanglement, the state of the quantum system being measured. The collapse and Born rules of standard quantum theory are thus recovered from a well-defined model.

In the CSL (Continuous Spontaneous Localisation) model of Pearle \cite{paper1}\cite{paper2}\cite{book}, a non-Hermitian Hamiltonian  is added to the usual Hamiltonian. This new Hamiltonian depends upon a random classical field and upon a set of completely commuting operators called the {\it collapse-generating operators}, whose joint eigenvectors are the possible collapse end results (in CSL these are chosen to be {\it smeared} mass-density operators).  The state vector smoothly evolves to one or another outcome (which outcome depends upon the specific realisation of the random field). There is no abrupt change of the state vector as in SL. The state vector norm is no longer conserved.  Instead, CSL provides a ``Probability Rule," specifying that the probability of each outcome is proportional to the state vector norm. This turns out to yield the Born rule. An alternative equivalent formulation is a normalised state vector evolution expressed as an It\^o stochastic differential equation that, however, is non-linear in the state vector.\footnote{Reference \cite{book} utilises the former formulation, but also provides chapters with a thorough introduction to  stochastic differential equations, followed by chapters describing the application of this formalism to CSL.} 

CSL is indistinguishable from standard quantum theory with current experimental capabilities. It predicts some effects which differ from standard quantum theory (see Ref.~\cite{book}, chapter 6, for an extensive discussion of these). 

Both SL and CSL are non-relativistic theories. An outstanding problem is to develop a satisfactory relativistic theory. This presents some technical problems. Localisation of a state invariably leads to spreading in momentum which results in an increase in the average energy. If, due to a particular random field, energy increases in one frame of reference, then we can boost to a frame where the boosted random field causes an  increase in energy that is arbitrarily large. Since all frames are equivalent, this implies that the energy increase is infinite in all frames. 

This infinite rate of energy increase in relativistic collapse was first pointed out in Ref.~\cite{infinite1} and discussed in  Ref.~\cite{infinite2}. This issue was amplified in a theorem by Myrvold~\cite{Myrvold}, who argued that, if a relativistic collapse model is to have a vacuum that is not excited by the random field, it must employ non-standard quantum degrees of freedom. Some relativistic models using non-standard degrees of freedom exist~\cite{Prel}\cite{Bed1} and serve as proof that relativistic collapse modes are possible in principle. 

In this paper we propose a new relativistic model, drawing inspiration from the postulated connection between collapse of the wave function and gravity. Several authors have made this connection. The approach of treating the collapse of the wave function as a physical process has been advocated by Roger Penrose~\cite{Pen} who has made a case that gravity is responsible for collapse and that this understanding could be an important step towards quantum gravity. An intriguing connection between collapse and Newtonian gravity has been made by Tilloy and Di\'osi \cite{TD}, and gravity-induced collapse is explored in, e.g.~\cite{Dio1}\cite{Dio2}\cite{Dio3}\cite{Fkar}. A recent proposal by Jonathan Oppenheim \cite{Opp} for a post-quantum theory of gravity in which quantum matter and classical gravity are unified employs mathematics which closely resembles the stochastic state dynamics of collapse models.

One might therefore draw on quantum gravity, with its own problems and attempted solutions, to see whether it might have any relevance to the problem of relativistic collapse models. One such issue is the problem of time, the name for a collection of conceptual problems related to the representation of time and quantisation. In
the canonical approach to quantum gravity, the Hamiltonian operator is found to be zero when acting on physical states (the Wheeler-DeWitt equation), thus the theory appears to be frozen in time. One response is to accept this fact and look for ways to recover time from the theory. An example of this approach is the work of Page and Wootters~\cite{PW} in which time is taken as a quantum degree of freedom, assigning to it its own Hilbert space. The passing of time is then understood as an entanglement between the time degree of freedom and the rest of the system. Other approaches also make use of a quantum time, see e.g.~\cite{QT} and references therein.

We follow this lead by treating time as a quantum operator. For simplicity, in this paper, we will only consider dynamics in one spatial dimension.  In addition to the position operators $\hat x_{i}$ for a collection of distinguishable particles labelled by $i$,\footnote{Our model concerns particles rather than quantum fields, a property it shares with  the relativistic collapse models considered by Tumulka \cite{RT1}\cite{RT2}.} we also introduce  time operators $\hat t_{i}$. Their conjugate operators are $\hat p_{i}, \hat E_{i}$, respectively, with non-vanishing commutators 
\begin{equation}\label{1}
[\hat x_{i}, \hat p_{j}]=i\delta_{ij},\hbox{      }[\hat t_{i}, \hat E_{j}]=-i\delta_{ij}.
\end{equation}
We interpret $E$ and $p$ as the energy and momentum of the particles. Note that we choose a negative sign for the $t$, $E$ commutator in order to conform with the usual definition of the energy operator $i\partial/\partial t$.
The bases we will use are $|x_{1}, x_{2},\ldots; t_{1},t_{2},\ldots\rangle$ or $|p_{1}, p_{2},\ldots; E_{1},E_{2},\ldots\rangle$.

We propose a model in which the collapse process takes place along a path in Hilbert space parametrised by the {\it evolutionary parameter} $s$. The state vector $|\psi, s\rangle$ is to obey a Schr\"odinger equation of the form
\[
\frac{d}{ds}|\psi, s\rangle=-i\hat H|\psi, s\rangle -\hat H'|\psi, s\rangle,
\]
where $\hat H' $ will depend on random functions of $s$ as well as collapse generating operators. In the course of this article we will examine how a picture of relativistic particles in spacetime emerges in which the particles exhibit localising behaviour consistent with Born rule probabilities. We will begin by considering one particle, then two, then many particles, in each case first considering the theory without collapse, then adding collapse. All these cases will be Poincar\'e invariant. 

\section{One Particle, No Collapse}

If we neglect the random part of the Hamiltonian, $\hat H'$, there are no collapses and the Schr\"odinger equation is
\begin{equation}\label{2}
\frac{d}{ds}|\psi, s\rangle=-i\hat H(\hat x,\hat t,\hat p, \hat E  )|\psi, s\rangle.
\end{equation}
We begin by addressing the question of how this state vector evolution is to be interpreted. After solving (\ref{2}), one has  the density $|\langle x,t|\psi,s\rangle|^{2}$, obeying 
\begin{align}
\int dxdt|\langle x,t|\psi,s\rangle|^{2}=1, \nonumber
\end{align}
which is the conditional probability, given $s$, that $x,t$ have particular values. However,  what one \textit{wants} to have is what one has in standard quantum theory, the conditional probability, given $t$, that the particle is at $x$. 

\subsection{Probability Considerations}

Suppose we start with an initial wave packet fairly well localised in spacetime, and envision it moving in spacetime as $s$ increases. One may think of this as the tracing out of a world-tube. We 
assign the same probability to each value of $s$. That is, if the initial state vector is defined at $s=-S/2$, and we consider the evolution until $s=S/2$, then the probability that the evolution parameter 
 lies between $s$ and  $s+ds$ is $ds/S$. Therefore, the joint probability density of $x,t$ throughout space-time is
 \begin{equation}\label{3}
{\cal P}(x,t)\equiv\frac{1}{S}\int_{-S/2}^{S/2}ds|\langle x,t|\psi,s\rangle|^{2}.
\end{equation}
\noindent We note that $\int dxdt{\cal P}(x,t)=1$ follows from Eq.~(\ref{2}). Since the probability of $t$  is ${\cal P}(t)\equiv \int dx {\cal P}(x,t) $, it follows that the conditional probability of $x$, given $t$, is
 \begin{equation}\label{4}
{\cal P}(x|t)=\frac{{\cal P}(x,t)}{{\cal P}(t)}.
\end{equation}
It is important to understand that this is not a probability density of the point $x$ at which we would find the particle if it were to be measured. Once we include collapses into our model, any measurements are fully described by the state evolution. Rather it is a density of where the particle {\it is} at time $t$.

\subsection{Relativistic Considerations}

\subsubsection{Poincar\'e Generators }

The Poincar\'e generators are space-translation $\hat p$, time-translation $\hat E$ and boost 
\[
\hat k=\hat x\hat E-\hat t\hat p,
\]
whose mutual non-vanishing commutation relations are, as usual,
\[
[\hat k,\hat p]=i\hat E,\quad [\hat k, \hat E]=i\hat p.
\]
We also note the commutation relations
\[
[\hat k,\hat x]=i\hat t, \quad[\hat k, \hat t]=i\hat x.
\]

Consider first the generation of boosts. The unitary transformation of boosts is readily calculated from the above commutation relations:
 \begin{eqnarray}
 \hat x'&=&e^{i\theta\hat k} \hat x e^{-i\theta\hat k}=\hat x\cosh\theta-\hat t\sinh\theta,\nonumber\\
  \hat t'&=&e^{i\theta\hat k} \hat t e^{-i\theta\hat k}=\hat t\cosh\theta-\hat x\sinh\theta,\nonumber\\
   \hat p'&=&e^{i\theta\hat k} \hat p e^{-i\theta\hat k}=\hat p\cosh\theta-\hat E\sinh\theta,\nonumber\\
    \hat E'&=&e^{i\theta\hat k} \hat E e^{-i\theta\hat k}=\hat E\cosh\theta-\hat p\sinh\theta.\nonumber
\end{eqnarray}
\noindent The relative velocity of the primed frame with respect to the unprimed frame is $w=\tanh\theta$, so $\cosh\theta\equiv\gamma=\frac{1}{\sqrt{1-w^{2}}}$, $\sinh\theta=w\gamma$. These, of course, are the 
correct Lorentz transformations and this confirms that we can regard $(\hat t,\hat x)$ and $(\hat E,\hat p)$ as operator-valued Lorentz vectors.

The space and time translations are
 \begin{eqnarray}
 \hat x'&=&e^{i[a\hat p-\tau \hat E]} \hat x e^{-i[a\hat p-\tau \hat E]}=\hat x+a, \nonumber\\
  \hat t'&=&e^{i[a\hat p-\tau \hat E]} \hat t e^{-i[a\hat p-\tau \hat E]}=\hat t+\tau, \nonumber\\
  \hat p'&=&e^{i[a\hat p-\tau \hat E]} \hat p e^{-i[a\hat p-\tau \hat E]}=\hat p, \nonumber\\
  \hat E'&=&e^{i[a\hat p-\tau \hat E]} \hat E e^{-i[a\hat p-\tau \hat E]}=\hat E. \nonumber
\end{eqnarray}

\subsubsection{Choice of Hamiltonian}

Applying the unitary boost operator to the Schr\"odinger equation, the boosted state vector $|\psi,s\rangle'=e^{i\theta\hat k} |\psi,s\rangle$ satisfies
 \begin{equation}\label{5}
\frac{d}{ds}|\psi, s\rangle'=-ie^{i\theta\hat k} \hat He^{-i\theta\hat k} |\psi, s\rangle'.
\end{equation}
For boost invariance, it is necessary that $|\psi, s\rangle'$ also be a possible evolution in the original frame, that is, a solution of the Schr\"odinger equation (\ref{2}), which requires  $\hat H'\equiv e^{i\theta\hat k} \hat He^{-i\theta\hat k} =\hat H$. This condition, 
that $\hat H$ be a Lorentz scalar means that it has to be a function of  Lorentz scalars. From our dynamical variables we can construct three Lorentz scalars, the scalar products of our two  
two-vectors $(\hat t, \hat x)$ and $(\hat E, \hat p)$:
\[
\hat x^{2}-\hat t^{2},\; \hat p^{2}-\hat E^{2},\; \hat x\hat p-\hat t\hat E.
\]

However,  the Lorentz scalars involving $\hat x, \hat t$ are not left invariant under the space-time translations. This means that the Hamiltonian can only depend upon $\hat p^{2}- \hat E^{2}$. 
For our illustrative example, we choose the Hamiltonian to be
 \begin{equation}\label{6}
\hat H\equiv\frac{1}{2m}[\hat p^{2}-\hat E^{2}].
\end{equation}
\noindent The symbol $m$  does not represent the mass of the particle. It is a constant having the dimension of mass so that the Hamiltonian has the dimension of energy.

\subsection{One-particle dynamics}
The joint probability density ${\cal P}(x,t)$ encodes all the information about where the particle can be found in space and time. Note that we can express ${\cal P}(x,t)$ by defining a density operator
 \begin{equation}\label{7}
\hat{\rho} = \frac{1}{S}\int_{-S/2}^{S/2} ds |\psi,s\rangle\langle \psi,s|,
\end{equation}
where we find
 \begin{equation}\label{8}
{\cal P}(x,t) = \langle x,t |\hat{\rho}| x,t\rangle.
\end{equation}
We highlight the fact that this description gives a picture of the particle in terms of a fixed state. We next consider how this picture describes the evolution of the position of the particle changing in time. 

As an example, we  consider a situation in which the wave function of the particle, in the position-time basis of a particular reference frame, is a Gaussian in both variables. Such a state is also Gaussian in the energy-momentum basis. The solution of the Schr\"odinger equation~(\ref{2}) in the energy-momentum basis is 
trivially 
\[
\langle p,E | \psi,s\rangle=C(p,E)e^{-i\frac{s}{2m}[p^{2}-E^{2}]},
\]
and we choose $C(p,E)$ as an appropriate normalised Gaussian:
 \begin{equation}\label{9}
\langle p,E | \psi,s\rangle = \left(\frac{2\sigma_x^{2}}{\pi}\right)^{1/4}e^{-\sigma_x^{2}(p-\bar{p})^2}e^{-i\bar{x} p }
\left(\frac{2\sigma_t^2}{\pi}\right)^{1/4}e^{-\sigma_t^{2}(E-\bar{E})^2}e^{i\bar{t} E }
e^{-i\frac{s}{2m}(p^{2}-E^{2})}.
\end{equation}
As we will see in Eqs.(\ref{10}) and (\ref{11}) below, this provides a separable state with respect to variables $x$ and $t$, in which each substate is Gaussian. 
The spreads of the Gaussians at evolution parameter value $s=0$  are $\sigma_x$ and $\sigma_t$ respectively, with larger values at $s=\pm S/2$. 
The central values of the packets at $s=0$  in $x,t$ space are $\bar{x}, \bar{t}$; the mean values of $E, p$ are $\bar{E}, \bar{p}$.

Taking the Fourier transform we find the state in the $x,t$ basis:
 \begin{align}\label{10}
\langle x,t |\psi,s\rangle =& \frac{1}{2\pi}\int dpdE e^{ipx -iEt} \langle p,E | \psi,s\rangle \nonumber\\
=& \left(\frac{\sigma_x^2}{2\pi\left(\sigma_x^2+is/2m\right)^2}\right)^{1/4}e^{-(\sigma_x \bar{p})^2}
e^{\frac{\left[i(x-\bar{x})+2\sigma_x^2 \bar{p}\right]^2}{4(\sigma_x^2 +is/2m)}} \nonumber\\
& \times 
\left(\frac{\sigma_t^2}{2\pi\left(\sigma_t^2-is/2m\right)^2}\right)^{1/4}e^{-(\sigma_t \bar{E})^2}
e^{\frac{\left[-i(t-\bar{t})+2\sigma_t^2 \bar{E}\right]^2}{4(\sigma_t^2 -is/2m)}},
\end{align}
and from this we find
 \begin{eqnarray}\label{11}
|\langle x,t |\psi,s\rangle |^2 =&
\frac{1}{\sqrt{2\pi[\sigma_x^2 + (s/2m\sigma_x)^2]}}e^{-\frac{[x-(\bar{x}+ s\bar{p}/m)]^2}{2[\sigma_x^2 + (s/2m\sigma_x)^2]}}
\nonumber\\
&\times 
\frac{1}{\sqrt{2\pi[\sigma_t^2 + (s/2m\sigma_t)^2]}}e^{-\frac{[t-(\bar{t}+s\bar{E}/m)]^2}{2[\sigma_t^2 + (s/2m\sigma_t)^2]}}.
\end{eqnarray}

Thus, as $s$  increases from $-S/2$ to $S/2$, the wave function traces out a world-tube, most dense along the world-line $(\bar{x} + s\bar{p}/m, \bar{t} + s\bar{E}/m)$, with a Gaussian profile  whose spread behaves as 
$\sqrt{\sigma_x^2+ (s/2m\sigma_x)^2}$ in the $x$ direction and  $\sqrt{\sigma_t^2+ (s/2m\sigma_t)^2}$ in the $t$ direction. If this were a classical world-line, for each value of $s$ there would be a point in the configuration space $x,t$. Here, each value of $s$ determines a 2 dimensional Gaussian region in the spacetime. The state at $s$ therefore connects the position of the particle with a given time. The spacetime probability density ${\cal P}(x,t)$ of equation (\ref{3}) is formed through a uniform weighting of these regions from $s=-S/2$ to $s=S/2$, thus correlating space and time for all values of $s$. 

In order to calculate ${\cal P}(x,t)$ we will assume that the parameter $m$ is sufficiently large that wave packet spreading can be ignored (later we will see how collapses can keep the state well localised thus preventing spreading). Provided that $S\ll m\sigma^{2}$, we can write
 \begin{eqnarray}\label{12}
{\cal P}(x,t) &\simeq& \frac{1}{S}\int^{S/2}_{-S/2}ds \frac{1}{\sqrt{2\pi\sigma_x^2}}e^{-\frac{[x-(\bar{x}+s\bar{p}/m)]^2}{2\sigma_x^2}}
\frac{1}{\sqrt{2\pi\sigma_t^2}}e^{-\frac{[t-(\bar{t}+s\bar{E}/m)]^2}{2\sigma_t^2}}\nonumber\\
&\simeq&\frac{1}{S}\frac{m}{\sqrt{2\pi(\sigma_x^2 \bar{E}^2 +\sigma_t^2 \bar{p}^2 )}}e^{-\frac{\left[\bar{E}(x-\bar{x}) - \bar{p}(t-\bar{t})\right]^2}{2(\sigma_x^2 \bar{E}^2 +\sigma_t^2 \bar{p}^2 )}},
\end{eqnarray}
where, in performing the integral, we have made the approximation that $S$ is large enough so that the integration limits can be assumed to be infinite\footnote{ It is clear from 
Eq.(\ref{3}) that $\int_{-\infty}^{\infty} dxdt{\cal P}(x,t)=\frac{1}{S}\int_{-S/2}^{S/2}1=1$. However, the approximation made in  (\ref{12}) invalidates that property [although it has negligible effect on ${\cal P}(x|t)$], as can be seen from 
 (\ref{13}), whose integral is infinite, not 1. One may view this approximation as effectively replacing the $-\infty,\infty$ limits of the $t$ integration by $-S\bar{E}/2m, S\bar{E}/2m$, so that the integral of   (\ref{13}) is 1. }. Hence, integrating over $x$,  we find
\begin{equation}\label{13}
{\cal P}(t) = \frac{m}{S\bar{E}},
\end{equation}
a uniform probability over possible values of $t$. Thus, from (\ref{4}), we obtain 
\begin{equation}\label{14}
{\cal P}(x|t) = \sqrt{\frac{\bar{E}^2}{2\pi(\sigma_x^2 \bar{E}^2 +\sigma_t^2 \bar{p}^2 )}}
e^{-\frac{[\bar{E}(x-\bar{x}) - \bar{p}(t-\bar{t})]^2}{2(\sigma_{x}^{2}\bar{E}^{2 } +\sigma_{t}^{2} \bar{p}^{2} )}}.
\end{equation}
The probability of $x$, conditional on $t$, thus describes a Gaussian wave packet of fixed width whose centre moves along a trajectory whose maximum density follows
\begin{equation}\label{15}
x = \bar{x} + \frac{\bar{p}}{\bar{E}}(t-\bar{t}\, ).
\end{equation}
This is a straight line path through spacetime and the velocity of the packet is $\bar{p}/\bar{E}$.

\section{One Particle, Including Collapse}\label{sec3}

Now we add dynamical wave function collapse to the single particle dynamics.  For a Poincar\'e invariant theory, not only must the Hamiltonian be a Lorentz scalar, but the
collapse generating operator $\hat A$ must be a Lorentz scalar as well. As we chose $\hat H$  by this criterion, so we choose
 \begin{equation}\label{16}
\hat A\equiv\hat p^{2}-\hat E^{2}.
\end{equation}
The state vector evolution can be written as an It\^o equation\footnote{This equation is derived in reference \cite{book}, where it appears as Eq. (16.9).  }
\begin{align}
d|\psi,s\rangle = \left\{-i\hat{H}ds - \frac{1}{2}\lambda (\hat A-\langle \hat A\rangle )^2 ds + ( \hat A-\langle \hat A\rangle )dB_s\right\}|\psi,s\rangle,
\label{17}
\end{align}
where $\langle \cdot \rangle = \langle \psi,s|\cdot |\psi,s\rangle$ and $B_s$ is a Brownian motion satisfying
\begin{align}
\mathbb{E}[dB_s] = 0, \quad (dB_s)^2 = \lambda ds,
\label{18}
\end{align}
where $\mathbb{E}$ denotes stochastic expectation. This produces a state evolution with fixed norm equal to 1. This means that with the definition of ${\cal P}(x,t)$ given in (\ref{3}), we still have $\int dxdt{\cal P}(x,t)=1$ and our definitions of probability remain valid.

\subsection{Density Matrix}
To discuss the behaviour of the complete ensemble of state vectors evolving under all possible realisations of the Brownian motion $B_s$ one looks at the density matrix defined by 
\begin{align}
\hat{\varrho}(s) =\mathbb{E}[ |\psi,s\rangle\langle \psi, s |],
\end{align}
whose evolution equation\footnote{Reference \cite{book}, Eq. (16.12). } is readily shown to follow from Eqs.(\ref{17}), (\ref{18}):
\begin{equation}\label{20}
\frac{d}{ds}\hat\varrho(s)=-i[\hat H, \hat\varrho(s)]-\frac{\lambda}{2}[\hat A,[\hat A,\hat\varrho(s)].
\end{equation}

A standard approach is to consider the solution of (\ref{20}) when $\hat H$ is set to zero to see how the collapse dynamics works without interference of the Hamiltonian dynamics. Then, in 
$\hat A$'s eigenbasis $ |a_{i}\rangle$ (chosen for simplicity to be countable, with all eigenvalues different):
\begin{equation}\label{21}
\langle a_{i}|\hat\varrho(s)|a_{j}\rangle=e^{-[s+S/2]\frac{\lambda}{2}[a_{i}-a_{j}]^{2}}\langle a_{i}|\hat\varrho(-S/2)|a_{j}\rangle.
\end{equation}
Equation (\ref{21}) encapsulates the collapse behaviour.  

When $s=S/2$, taking $S\lambda>>1$ , the right-hand side asymptotically essentially vanishes unless $a_{i}=a_{j}$. Thus, in the ensemble of state vectors evolving under the various possible realised Brownian motions, all asymptotically approach one or another eigenstate. For if one state vector in the ensemble was not an eigenstate but rather a superposition of eigenstates, it could have contributed a non-zero off-diagonal element. 
  
Moreover, the asymptotic ensemble obeys the Born rule. For, if 
\begin{align} 
\hat\varrho(-S/2)=\sum_{ij}c_{i}c_{j}^{*}|a_{i}\rangle\langle a_{j}|,\nonumber
\end{align} 
i.e.~a pure state with amplitude $c_i$ associated to each eigenstate $|a_i\rangle$, then according to (\ref{21}), the probability associated to the $i$th outcome is $\langle a_{i}|\hat\varrho(S/2)|a_{i}\rangle=|c_{i}|^{2}$, the probabilities of the various states in the initial superposition.

In our example, we need not set $\hat H$ equal to zero since it commutes with $\hat A$ and therefore does not interfere with the collapse behaviour. The solution of (\ref{20}) is, in the momentum-energy basis,
\begin{align}\label{22}
\langle p,E|\hat\varrho(s)|p',E'\rangle=&e^{ -i\frac{1}{2m}[s+S/2][(p^{2}-E^{2})-(p'^{2} -E'^{2})]}e^{-[s+S/2]\frac{\lambda}{2}[[(p^{2}-E^{2})-(p'^{2} -E'^{2})]^{2}} \nonumber\\
&\times \langle p,E|\hat\varrho(-S/2)|p',E'\rangle.
\end{align}
  
 Taking $S\lambda>>1$ , the right hand side essentially vanishes unless $p^{2}-E^{2}=p'^{2} -E'^{2}$. Changing the labelling 
  of the momentum-energy basis, we have that the non-vanishing asymptotic density matrix elements are just the  elements diagonal in $p^{2}-E^{2}$:
\begin{equation}\label{23}
\langle p^{2}-E^{2}, p+E|\hat\varrho(S/2)|p^{2} -E^{2}, p'+E'\rangle= \langle p^{2}-E^{2}, p+E|\hat\varrho(-S/2)|p^{2} -E^{2}, p'+E'\rangle.
\end{equation}  

Now we show that the asymptotic states satisfy the Klein-Gordon equation. For, we have shown that the asymptotic (collapsed) states are eigenstates of $\hat E^2 -\hat p^2$ with 
eigenvalues $E^{2}-p^{2}\equiv\mu^2$ (we assume the initial state is one which is a superposition of states of non-negative $\mu^2$). Changing the labelling of the momentum-energy basis once again, we represent such a collapsed state as a superposition of states with a definite value of $\mu$ and arbitrary dependence on $p$:
\begin{equation}\label{24}
|\psi,S/2\rangle = N\int dp\phi(p)|\omega_{p}, p\rangle,
\end{equation}
where $\omega_{p} = (p^{2} + \mu^{2})^{1/2}$, $\langle E', p'|\omega_{p}, p\rangle=\delta(E'-\omega_{p} )\delta(p'-p)$, $\phi(p)$ is some arbitrary function and $N$ is a normalisation factor. In the position-time basis the particle wave function is 
\begin{eqnarray}\label{25}
\langle x,t|\psi,S/2\rangle&=& N\int dp\phi(p)\langle x,t|\omega_p, p\rangle\nonumber\\
&=&N\int dp\phi(p)\int dp'dE'\langle x,t|E', p'\rangle\langle E', p'|\omega_p, p\rangle\nonumber\\ 
&=&\frac{N}{2\pi}\int dp\phi(p)\int dp'dE'e^{ip'x}e^{-iE't}\delta(E'-\omega_{p} )\delta(p'-p)\nonumber\\ 
&=&\frac{N}{2\pi}\int dp\phi(p)e^{ipx}e^{-i\omega_{p} t}.
\end{eqnarray}

The dynamics has therefore resulted in collapsed states,  each state in the ensemble satisfying the Klein Gordon equation:
\begin{equation}\label{26}
\Bigg(\frac{\partial^{2}}{\partial t^{2}} -\frac{\partial^{2}}{\partial x^{2}}+ \mu^2\Bigg)\langle x,t |\psi, S/2\rangle = 0.
\end{equation}
This illustrates the relativistic invariance of our model (a solution of the Klein-Gordon equation in one frame is a solution in all frames).

Equation (\ref{26}) describes a particle with positive mass squared provided that the eigenvalue $\mu^2$ of $-\hat A$ remains positive. To test this we calculate the process for $\langle \hat A\rangle$. Applying the It\^o rules to  equations (\ref{17}) and (\ref{18}) we find (note: $\langle \hat A^{2}\rangle - \langle \hat A\rangle^2 =\langle (\hat A - \langle \hat A\rangle)^2 \rangle$):
\begin{align}\label{27}
d\langle \hat A \rangle  = 2\langle (\hat A - \langle \hat A\rangle)^2 \rangle dB_s.
\end{align}
Since  $\langle \hat A \rangle=-\mu^2$, the particles which are the end states of collapse need not have positive mass squared, since $dB_s$ can be positive as well as negative. 
This equation implies that, if the quantum variance in the state of $\hat A$ is not $0$, then there can be negative mass squared, i.e., tachyonic behaviour. We accept that the model allows tachyons, whilst noting that  they could be effectively prevented by having an initial state composed of only eigenstates of $\hat A$ with negative eigenvalues and having a sufficiently strong collapse mechanism in order that the quantum variance in $\hat A$ rapidly tends to zero such that stochastic movements in $\langle \hat A \rangle$ are suppressed. This would quickly result in a positive value of $\mu^2$.

The considerations of this section also confirm that we can regard $\hat E$ as corresponding to the relativistic energy of the particle. The process for the quantum expectation of $\hat E$ is similarly found to be 
\begin{align}\label{P0}
d\langle \hat E \rangle = \langle \{(\hat E-\langle \hat E\rangle),(\hat A - \langle \hat A \rangle )\}\rangle dB_s
\end{align}
($ \{. , .\}$ is the anti-commutator symbol).  This shows that energy is conserved in stochastic expectation for this model. 

\subsection{Further considerations}

The concept of the state developing as the parameter $s$ increases does not necessarily equate to a state development with increasing $t$. In order to understand how the dynamics in $s$ leads to development in time, let us ask how the quantum expectation of the time operator $\hat t$ for a state $|\psi,s\rangle$ changes with $s$. We find using (\ref{17}) and (\ref{18}) that
\begin{align}
d\langle \hat t \rangle = \frac{\langle \hat E \rangle}{m}ds 
+ \langle \{(\hat t - \langle \hat t \rangle )  , (\hat A - \langle \hat A \rangle )\}\rangle dB_s.
\end{align}
This equation shows that as $s$ advances there is a drift in the expected time given by the first term on the right side. This drift is positive (i.e.~time increases in expectation with increasing $s$) provided that the energy $\langle \hat E \rangle$ is positive. (Moreover, we see that $\langle \hat E \rangle/m$ represents the rate of flow of time with $s$.) From equation (\ref{P0}) we see that, despite the energy being conserved in stochastic expectation, it can vary stochastically for a given realisation of the Brownian motion $B_s$. Again these stochastic movements are suppressed if the collapse mechanism is strong.

Therefore, the model does not strictly prevent a situation in which the time  of the particle state decreases as $s$ increases. We point out that in this case, an observer moving in a forward direction of time would see a sequence of collapses and, although this sequence is generated by a backwards-in-time collapse process, as shown in Ref.~\cite{Bed3} where the time reversal symmetry of collapse models is examined, the sequence would appear as if generated by a forward-in-time collapse process. 

Provided that energy remains positive, time will increase on average as $s$ increases and we can treat $s$ as a proxy for time. 

To conclude, in this one particle model we have shown that collapse leads to behaviour which closely resembles the relativistic motion of a particle with well defined mass $\mu$. What the collapse behaviour of the  model does not do is spatially localise the state. This can be introduced by looking at a situation with more particles. We look at this in the next section.

\section{Two Particles, No Collapse}
\label{sec:2PNC}

With two distinguishable particles, there are eight operators, the position-time operators $\hat x_{1}, \hat t_{1},\hat x_{2},\hat t_{2}$ and their conjugate operators $\hat p_{1}, \hat E_{1},p_{2}, \hat E_{2}$, obeying the commutation relations~(\ref{1}).

We choose the Hamiltonian
\begin{eqnarray}\label{30}
\hat H&=&\frac{1}{2m_{1}}[\hat p_{1}^{2}-\hat E_{1}^{2}] +\frac{1}{2m_{2}}[\hat p_{2}^{2}-\hat E_{2}^{2}],
\end{eqnarray}
\noindent where $m_1$, $m_2$ are constants with dimensions of mass.
We see from (\ref{30}) that $\hat H$ is composed of two Poincar\'e scalars, so $\hat H$ is a Poincar\'e scalar, and the Schr\"odinger equation is Poincar\'e invariant.

As in section IIC, as an example, we choose a solution of Schr\"odinger's equation that is a Gaussian in all variables. Thus, we augment Eq.(\ref{9})
 \begin{align}\label{31}
\langle p_1, E_1,p_2,E_2 | \psi,s\rangle =& \left(\frac{2\sigma_{x_1}^{2}}{\pi}\right)^{1/4}e^{-\sigma_{x_1}^{2}(p_1-\bar{p}_1)^2}e^{-i\bar{x}_{1} p_1 }\nonumber\\
&\times\left(\frac{2\sigma_{t_1}^2}{\pi}\right)^{1/4}e^{-\sigma_{t_1}^{2}(E_1-\bar{E}_1)^2}e^{i\bar{t}_{1} E_1 }
e^{-i\frac{s}{2m_1}(p_1^{2}-E_1^{2})}\nonumber\\
&\times  \left(\frac{2\sigma_{x_2}^{2}}{\pi}\right)^{1/4}e^{-\sigma_{x_2}^{2}(p_2-\bar{p}_2)^2}e^{-i\bar{x}_{2} p_2 }
\nonumber\\
&\times\left(\frac{2\sigma_{t_2}^2}{\pi}\right)^{1/4}e^{-\sigma_{t_2}^{2}(E_2-\bar{E}_2)^2}e^{i\bar{t}_{2} E_2 }
e^{-i\frac{s}{2m_2}(p_2^{2}-E_2^{2})},
\end{align}
take the Fourier transform as in (\ref{10}), and end up with the equivalent of (\ref{11}):
 \begin{align}\label{32}
|\langle x_1,t_1,x_2,t_2 |\psi,s\rangle |^2 =&
\frac{1}{\sqrt{2\pi[\sigma_{x_1}^2 + (s/2m_1\sigma_{x_1})^2]}}e^{-\frac{[x_1-(\bar{x}_1+s\bar{p}_1/m_1)]^2}{2[\sigma_{x_1}^2 + (s/2m_1\sigma_{x_1})^2]}}\nonumber\\
&\times\frac{1}{\sqrt{2\pi[\sigma_{t_1}^2 + (s/2m_1\sigma_{t_1})^2]}}e^{-\frac{[t_1-(\bar{t}_1+s\bar{E}_1/m_1)]^2}{2[\sigma_{t_1}^2 + (s/2m_1\sigma_{t_1})^2]}}\nonumber\\
&\times\frac{1}{\sqrt{2\pi[\sigma_{x_2}^2 + (s/2m_2\sigma_{x_2})^2]}}e^{-\frac{[x_2-(\bar{x}_2+ s\bar{p}_2/m_2)]^2}{2[\sigma_{x_2}^2 + (s/2m_2\sigma_{x_2})^2]}}\nonumber\\
&\times\frac{1}{\sqrt{2\pi[\sigma_{t_2}^2 + (s/2m_2\sigma_{t_2})^2]}}e^{-\frac{[t_2-(\bar{t}_2+s\bar{E}_2/m_2)]^2}{2[\sigma_{t_2}^2 + (s/2m_2\sigma_{t_2})^2]}}.
\end{align}

We define the density associated with multiple particles by
\begin{align}\label{33}
{\cal P}(x_1,t_1; x_2,t_2; \ldots) = {\cal P}(x_1,t_1){\cal P}(x_2,t_2)\cdots,
\end{align}
where the marginal densities are given by
\begin{align}\label{34}
{\cal P}(x_i, t_i) = \frac{1}{S}\int_{-S/2}^{S/2} ds \langle \psi, s |\left(|x_i,t_i\rangle\langle x_i,t_i|\otimes \mathbb{1}_{\neq i}\right)|\psi,s\rangle
\end{align}
(the operator $\mathbb{1}_{\neq i}$ is the unit operator applied to all degrees of freedom other than $x_i$, $t_i$).
We are treating the overall density as though each particle density is independent. This doesn’t mean that there cannot be interesting interactions and entanglements. This all happens at the state level. This probability density is a way of mapping that information into spacetime.

The density for $x_1, x_2, \ldots$ conditional on $t_1, t_2, \ldots$ is then given by the formula for conditional probability
\begin{align}\label{35}
{\cal P}(x_1 x_2  \ldots |t_1, t_2, \ldots) = \frac{{\cal P}(x_1,t_1; x_2,t_2; \ldots)}{{\cal P}(t_1, t_2,\ldots)},
\end{align}
where
\begin{align}\label{36}
{\cal P}(t_1, t_2, \ldots )  = \int dx_1 dx_2 \cdots {\cal P}(x_1,t_1; x_2,t_2; \ldots).
\end{align}
This means that
\begin{align}\label{37}
{\cal P}(x_1 x_2  \ldots |t_1, t_2, \ldots)  = \frac{{\cal P}(x_1,t_1)}{{\cal P}(t_1)}\frac{{\cal P}(x_2,t_2)}{{\cal P}(t_2)}\cdots
 = {\cal P}(x_1|t_1){\cal P}(x_2|t_2)\cdots,
\end{align}
where
\begin{align}\label{38}{\cal P}(t_i) = \int dx_i {\cal P}(x_i, t_i).
\end{align}

For the state defined by (\ref{31}) we thus have for particle 1
\begin{align}\label{39}
{\cal P}(x_1,t_1) = \frac{1}{S}\int ds & dx_2dt_2 |\langle x_1,t_1,x_2,t_2 |\psi,s\rangle |^2 \nonumber\\
=\frac{1}{S}\int ds & \frac{1}{\sqrt{2\pi[\sigma_{x_1}^2 + (s/2m_1\sigma_{x_1})^2]}}e^{-\frac{[x_1-(\bar{x}_1+ s\bar{p}_1/m_1)]^2}{2[\sigma_{x_1}^2 + (s/2m_1\sigma_{x_1})^2]}}\nonumber\\
&\times\frac{1}{\sqrt{2\pi[\sigma_{t_1}^2 + (s/2m_1\sigma_{t_1})^2]}}e^{-\frac{[t_1-(\bar{t}_1+s\bar{E}_1/m_1)]^2}{2[\sigma_{t_1}^2 + (s/2m_1\sigma_{t_1})^2]}}.
\end{align}

To do the $s$ integral, as in Section IIC, we assume that the parameters $m_i$ are sufficiently large to allow the approximation
\begin{align}\label{40}
{\cal P}(x_1,t_1) 
&=\frac{1}{S}\int ds  \frac{1}{\sqrt{2\pi \sigma_{x_1}^2}}e^{-\frac{[x_1-(\bar{x}_1+ s\bar{p}_1/m_1)]^2}{2\sigma_{x_1}^2}}
\frac{1}{\sqrt{2\pi\sigma_{t_1}^2}}e^{-\frac{[t_1-(\bar{t}_1+s\bar{E}_1/m_1)]^2}{2\sigma_{t_1}^2 }} \nonumber\\
&\simeq \frac{1}{S}\frac{m_1}{\sqrt{2\pi(\sigma_{x_1}^2 \bar{E}_1^2+ \sigma_{t_1}^2\bar{p}_1^2)}} 
e^{-\frac{[\bar{E}_1(x_1-\bar{x}_1)-\bar{p}_1(t_1-\bar{t_1})]^2} {2(\sigma_{x_1}^2 \bar{E}_1^2+ \sigma_{t_1}^2\bar{p}_1^2)}}.
\end{align}
From this we can calculate
\begin{align}\label{41}
{\cal P}(t_1) =\frac{m_1}{S\bar{E}_1}, 
\end{align}
and 
\begin{align} \label{px1t1}
{\cal P}(x_1|t_1) = \sqrt{\frac{\bar{E}^2_1}{{2\pi(\sigma_{x_1}^2 \bar{E}_1^2+ \sigma_{t_1}^2\bar{p}_1^2)}} }
e^{-\frac{[\bar{E}_1(x_1-\bar{x}_1)-\bar{p}_1(t_1-\bar{t_1})]^2} {2(\sigma_{x_1}^2 \bar{E}_1^2+ \sigma_{t_1}^2\bar{p}_1^2)}}.
\end{align}

This means that the probability density is focussed about a path through spacetime determined by
\begin{align}\label{43}
x_1 = \bar{x}_1 + \frac{\bar{p}_1}{\bar{E}_1}(t_1 - \bar{t}_1).
\end{align}
Particle 1 follows a straight line path through spacetime with velocity $\bar{p}_1/\bar{E}_1$.
The same calculation for particle 2 leads to the same result as (\ref{px1t1}) with 2 in place of 1.

\section{Two Particles, Including Collapse}

For Poincar\'e invariance, the collapse-generating operators must be Poincar\'e invariant scalars. With two particles there are several of these including $\hat p_1^{2}-\hat E_1^{2}$, $\hat p_2^{2}-\hat E_2^{2}$, and 
\begin{align}\label{44}
 \Delta \hat x^2 - \Delta \hat t^2,
\end{align}
where $\Delta \hat x = \hat x_1 - \hat x_2$, and $\Delta \hat t = \hat t_1 - \hat t_2$.
This set of operators are not completely commuting but this doesn't prevent us from considering a collapse model in which the collapse generating operators are this set of scalars. It simply makes the solutions harder to find since there is no common eigenbasis. With $\hat H$ given by equation (\ref{30}), a general It\^o state evolution including collapse is 
\begin{align}\label{gensde}
d|\psi,s\rangle  = \left\{ - i\hat H ds
-\frac{1}{2}(\hat{\bf A} - \langle \hat{\bf A}\rangle)\cdot {\bf \Lambda} \cdot (\hat{\bf A} - \langle \hat{\bf A}\rangle)ds
+(\hat{\bf A} - \langle \hat{\bf A}\rangle)\cdot d{\bf B}_s
\right\}|\psi,s\rangle,
\end{align}
where $\hat{\bf A} = (\hat A_1, \hat A_2, \ldots)$, ${\bf B}_s = (B_{1,s},B_{2,s}, \ldots)$ and
\begin{align}\label{46}
{\bf \Lambda} = {\rm diag}(\lambda_1, \lambda_2,\ldots),
\end{align}
where $\lambda_i$ are the collapse strengths which can be different for each collapse generating operator $\hat A_i$. The components of the Brownian motion vector satisfy
\begin{align}\label{47}
\mathbb{E}[dB_{i}] = 0, \quad dB_{i}dB_{j} = \lambda_i\delta_{ij}ds.
\end{align}

We choose a specific model with 3 collapse generating operators
\begin{align}\label{3cgo}
\hat A_1 &= (\hat p_1^{2}-\hat E_1^{2}) ,\nonumber\\
\hat A_2 &= (\hat p_2^{2}-\hat E_2^{2}) ,\nonumber\\
\hat A_3 &= ( \Delta \hat x^2 - \Delta \hat t^2).
\end{align}
Since the collapse generating operators are not mutually commuting, they are in some sense competing and the collapse process will force the state to have minimal variance in each $\hat A_i$ subject to uncertainty principle constraints. By varying the strengths $\lambda_i$, the relative optimal variance of each $\hat A_i$ will vary.

Since we have already examined the effect of collapse generating operators like $\hat A_1$ and $\hat A_2$ in section \ref{sec3} we shall ignore them here, noting only that their effect is to drive the separate particle states into states of well-defined mass.

We therefore focus on the collapse generating operator $\hat A_3$ (henceforth simply $\hat A$) and consider the simplified collapse dynamics
\begin{align}\label{49}
d|\psi,s\rangle  = \left\{ -i\hat H ds
-\frac{1}{2}\lambda (\hat{A} - \langle \hat{A}\rangle)^2ds
+(\hat{A} - \langle \hat{A}\rangle) d{B}_s
\right\}|\psi,s\rangle,
\end{align}
with 
\begin{align}\label{50}
\hat A = \Delta \hat x^2 - \Delta \hat t^2,
\end{align}
and 
\begin{align}\label{51}
\mathbb{E}[dB_{s}] = 0, \quad (dB_{s})^2 = \lambda ds.
\end{align}
As mentioned, as is usual in evaluating collapse behavior, we henceforth set $\hat H=0$.

We can represent the state in the $|x_1,t_1,x_2,t_2\rangle$ basis. The solution to equation (\ref{20}) represented in this basis is 
\begin{align}\label{52}
\langle x_1,t_1,x_2,t_2|\hat\varrho(s)|x'_1,t'_1,x'_2,t'_2\rangle=&
e^{-[s+S/2]\frac{\lambda}{2}[(\Delta \hat x^2 - \Delta \hat t^2)-(\Delta \hat x'^2 - \Delta \hat t'^2)]^{2}}        \nonumber\\
&\times
 \langle x_1,t_1,x_2,t_2|\hat\varrho(-S/2)|x'_1,t'_1,x'_2,t'_2\rangle.
\end{align}
In order to examine the collapse behaviour let us consider a special state such that $\Delta \hat t $ can be taken as well localised with value close to zero in some frame of reference. [We note that this assumption is unstable since if $\Delta \hat t$ is well localised then by the uncertainty principle there must be a spread in the conjugate variable $\Delta \hat E = (\hat E_1 - \hat E_2)$ which means that there must be contributions to the state where the energy of particle 1 is different from that of particle 2. This effectively means that time will flow at a different rate (with respect to $s$) for particle 1 than for particle 2 and so a spread in the state of $\Delta \hat t$ will develop.
However, we can drop the assumption of localised relative time in the next section when we examine the case of $N$ particles.]
For such a state we have
\begin{align}\label{53}
\langle x_1,t_1,x_2,t_2 |\hat\varrho(S/2)|x'_1,t'_1,x'_2,t'_2 \rangle = &
e^{-S\frac{\lambda}{2}[\Delta x^{2}-\Delta x'^{2}]^{2}}\nonumber \\
&\times \langle x_1,t_1,x_2,t_2|\hat\varrho(-S/2)|x'_1,t'_1,x'_2,t'_2\rangle.
\end{align}
For sufficiently large $S\lambda$, we see the familiar signature of collapse to eigenstates of $\Delta \hat x^{2}$, i.e.~asymptotic vanishing of the off-diagonal elements of the density matrix, with satisfaction of the Born rule. Thus we can obtain spatially localised collapse.

\subsection{An example of a state for which collapse does not occur}

The hallmark of a collapse theory is that a state describing a superposition of two states of particles in two places collapses to one or another of these states with Born rule probabilities. Consider the reduced density matrix of only spatial states obtained by tracing out the time degrees of freedom described by the basis states $| t_1, t_2\rangle$,
\begin{align}\label{54}
\hat \varrho_{\rm space} = {\rm Tr}_{\rm time}\hat \varrho.
\end{align}
Suppose that the initial space state is constructed from states $|\phi_L\rangle$ and $|\phi_R\rangle$ with wavefunctions $\langle x_1, x_2|\phi_{L}\rangle$ describing both particles 1 and 2 being localised within a distance $a$ about position $x_i = L$, and similarly for the $R$-state (where the two particles are both localised within a distance $a$ about $x_i = R$), with $a \ll |R-L|$. We suppose that the initial state is an equal superposition of $L$ and $R$ states
\begin{align}\label{55}
\hat \varrho_{\rm space} (-S/2) = \frac{1}{2}[|\phi_L\rangle + |\phi_R\rangle]
[\langle\phi_L | + \langle\phi_R |].
\end{align}

The asymptotic reduced density matrix, by  (\ref{53}), is therefore
\begin{eqnarray}\label{56}
\langle x_{1}, x_{2}|\hat \varrho_{\rm space} (S/2)| x'_{1}, x'_{2}\rangle&=&e^{-S\frac{\lambda}{2}[\Delta x^{2}-\Delta x'^{2}]^{2}} \langle x_{1}, x_{2}|\hat \varrho_{\rm space} (-S/2)| x'_{1}, x'_{2}\rangle.
\end{eqnarray}
The diagonal matrix element associated to the $L$ state is 
\begin{align}\label{57}
\langle \phi_{L}|\hat \varrho_{\rm space} (S/2)|\phi_{L}\rangle=&\int dx_{1}d x_{2}dx'_{1} dx'_{2}
\langle \phi_{L}|x_{1}, x_{2}\rangle \langle x'_{1}, x'_{2}|\phi_{L}\rangle
              e^{-S\frac{\lambda}{2}[(x_{1}-x_{2})^{2}- (x'_{1}-x'_{2})^{2}]^{2}} \nonumber\\
              &\;\times \frac{1}{2}[\langle x_{1}, x_{2}|\phi_{L}\rangle+\langle x_{1}, x_{2}|\phi_{R}\rangle]
[\langle \phi_{L}|x'_{1},x'_{2}\rangle+\langle \phi_{R}|x'_{1}, x'_{2}\rangle] \nonumber\\ 
=&\frac{1}{2}\int dx_{1}d x_{2}dx'_{1} dx'_{2}
|\langle \phi_{L}|x_{1}, x_2\rangle |^{2}|\langle \phi_{L}|x'_{1}, x'_2\rangle |^{2}
              e^{-S\frac{\lambda}{2}[(x_{1}-x_{2})^{2}- (x'_{1}-x'_{2})^{2}]^{2}}.
\end{align}
If, for simplicity, we let $a$ approach 0 so  $|\langle \phi_{L}|x_1,x_2\rangle |^{2} =\delta(x_1-L)\delta(x_2-L)$ and the exponential equals 1, then $\langle \phi_{L}|\hat \varrho_{\rm space} (S/2)|\phi_{L}\rangle=1/2$.

However, the off-diagonal matrix element does not vanish: 
\begin{align}\label{58}
\langle \phi_{L}|\hat \varrho_{\rm space} (S/2)|\phi_{R}\rangle =&\int dx_{1}d x_{2}dx'_{1} dx'_{2}
\langle \phi_{L}|x_{1},x_{2}\rangle \langle x'_{1},x'_{2}|\phi_{R}\rangle
              e^{-S\frac{\lambda}{2}[(x_{1}-x_{2})^{2}- (x'_{1}-x'_{2})^{2}]^{2}} \nonumber\\
              &\;\times\frac{1}{2}\langle x_{1}, x_{2}|\phi_{L}\rangle
\langle \phi_{R}|x'_{1},x'_{2}\rangle \nonumber\\ 
=&\frac{1}{2}\int dx_{1}d x_{2}dx'_{1} dx'_{2} e^{-S\frac{\lambda}{2}[(x_{1}-x_{2})^{2}- (x'_{1}-x'_{2})^{2}]^{2}}\nonumber \\
&\;\times \delta(x_{1}-L)\delta(x_{2}-L)\delta(x'_{1}-R)\delta(x'_{2}-R)\nonumber\\
 =&\frac{1}{2}.
\end{align}
Therefore, the superposition has not collapsed.

It is easy to see why collapse does not happen in this example. With the assumptions that we have made, both the $|\phi_L\rangle $ and $|\phi_R\rangle$ states are eigenstates of $\Delta \hat x$ with the same eigenvalue
\begin{align}\label{59}
\Delta \hat x | \phi_{L/R}\rangle = (\hat x_1 - \hat x_2) | \phi_{L/R}\rangle = 0.
\end{align}
These two states are therefore not differentiated by the collapse process. If we imagine the world as made up of only these two particles, then any $\Delta x$-states that are the same can be viewed as the same configuration of the two particles. The collapse process does not treat such states as different. From a configuration perspective, the two states $|\phi_L\rangle$ and $|\phi_R\rangle$ are essentially the same state of both particles being close together in space.

In order to see collapse in this model we should construct an initial state of different $\Delta x$-configurations. 

\subsection{An example of a state for which collapse occurs}

Let us take the initial pure density matrix for the two particles to be a superposition of only particle 1 states. Suppose that the initial state is now constructed from states $|\chi_L\rangle $ and $|\chi_R\rangle$. The wavefunction $\langle x_1, x_2|\chi_L\rangle$ describes particle 1 being localised about a position $x_1 = L $ and particle 2 being localised about a different position $x_2 = C$; the wavefunction $\langle x_1, x_2|\chi_R\rangle$ describes particle 1 being localised about position $x_1 = R $ with particle 2 being localised about the same position $x_2 = C$, where $|L-C| \neq |R-C|$ and the localisation scale $a\ll |L-C| , |R-C|$. Take the initial reduced state to be an equal superposition
\begin{align}\label{60}
\hat \varrho_{\rm space} (-S/2) = \frac{1}{2}[|\chi_L\rangle + |\chi_R\rangle]
[\langle\chi_L | + \langle\chi_R |].
\end{align}
For these states we have
\begin{align}\label{61}
\Delta \hat x | \chi_{L}\rangle = (L-C) | \chi_{L}\rangle ; \quad \Delta \hat x | \chi_{R}\rangle = (R-C) | \chi_{R}\rangle ,
\end{align}
demonstrating that they represent different configurations provided that $|L-C| \neq |R-C|$.

The asymptotic density matrix is given in the $x_1, x_2$ basis by
\begin{align}\label{62}
\langle x_{1}, & x_{2}|\hat \varrho_{\rm space} (S/2)| x'_{1}, x'_{2}\rangle \nonumber\\
&=e^{-S\frac{\lambda}{2}[\Delta x^{2}- \Delta x'^{2}]^{2}}\frac{1}{2}[\langle x_{1},x_2|\chi_{L}\rangle+\langle x_{1},x_2|\chi_{R}\rangle]
[\langle \chi_{L}|x'_{1},x'_2\rangle +\langle \chi_{R}|x'_{1},x'_2\rangle]\nonumber\\
&= \frac{1}{2}\left\{[e^{-S\frac{\lambda}{2}[(L-C)^{2}- (L-C)^{2}]^{2}}\langle x_{1},x_2|\chi_{L}\rangle\langle \chi_{L}|x'_{1},x'_2\rangle\right. \nonumber\\
 &\quad\quad + e^{-S\frac{\lambda}{2}[(R-C)^{2}- (R-C)^{2}]^{2}}\langle x_{1},x_2|\chi_{R}\rangle\langle \chi_{R}|x'_{1},x'_2\rangle\nonumber\\
&\quad\quad+\left.e^{-S\frac{\lambda}{2}[(L-C)^{2}- (R-C)^{2}]^{2}}\left[\langle x_{1},x_2|\chi_{L}\rangle\langle \chi_{R}|x'_{1},x'_2\rangle+\langle x_{1},x_2|\chi_{R}\rangle\langle \chi_{L}|x'_{1},x'_2\rangle\right]
 \right\}\nonumber\\
 &= \frac{1}{2}\left[\langle x_{1},x_2|\chi_{L}\rangle\langle \chi_{L}|x'_{1},x'_2\rangle+\langle x_{1},x_2|\chi_{R}\rangle\langle \chi_{R}|x'_{1},x'_2\rangle\right].
 \end{align}
where we have assumed that $|L-C|\neq|R-C|$. This demonstrates that the model achieves collapse of a superposition of different spatial configurations of particles.

\subsection{Energy Process}
For the model defined by equations (\ref{gensde})-(\ref{3cgo}) we can calculate the stochastic process for the quantum expectation of the particle energy $\langle \hat E_i\rangle$. Using the usual rules of It\^o calculus we find
\begin{align}\label{econ1}
d\langle \hat E_i\rangle = \sum_{j=1}^3\langle\{(\hat E_i- \langle \hat E_i\rangle, (\hat A_j - \langle \hat A_j \rangle)\}\rangle dB_j . 
\end{align}
This means that the quantum expectation of the energy for each particle is conserved in stochastic expectation:
\begin{align}\label{econ2}
d\mathbb{E}[\langle \hat E_i\rangle] = 0.
\end{align}
The energy of each particle is conserved on average. The quantum expectation of the energy will vary stochastically for a given realisation of the Brownian motion in proportion to the covariance between $\hat E_i$ and $\hat A_j$

\subsection{Processes for particle mass and invariant separation}
The stochastic processes for the quantum expectations $\langle \hat A_i \rangle$ with $\hat A_i$ defined by (\ref{3cgo}) are found to be
\begin{align}
d\langle \hat A_1 \rangle &=  \sum_{i = 1}^3 \langle \{ \hat A_1,(\hat A_i - \langle \hat A_i \rangle) \}\rangle dB_i +4\lambda_3\langle \hat A_3 \rangle ds, \label{A1prc}\\
d\langle \hat A_2 \rangle &=  \sum_{i = 1}^3 \langle \{ \hat A_2,(\hat A_i - \langle \hat A_i \rangle) \}\rangle dB_i +4\lambda_3\langle \hat A_3 \rangle ds, \label{A2prc}\\
d\langle \hat A_3 \rangle &= i\langle [\hat H, \hat A_3 \rangle ds + \sum_{i = 1}^3 \langle \{ \hat A_3,(\hat A_i - \langle \hat A_i \rangle) \}\rangle dB_i+4\lambda_1\langle \hat A_1 \rangle ds + 4\lambda_2\langle \hat A_2 \rangle ds. \label{A3prc}
\end{align}
We interpret $A_1$ and $A_2$ as the negative of the mass squared of particles 1 and 2 respectively; $A_3$ represents the invariant separation of the two particles in spacetime. The stochastic terms (terms proportional to $dB_i$) in each of these equations are proportional to the quantum covariances between the different operators $\hat A_i$. The first term in equation (\ref{A3prc}) represents the changes that occur in invariant separation of particles due to the motion of the particles determined by the Hamiltonian (see Section \ref{sec:2PNC}). The remaining terms represent additional drifts in the $\langle \hat A_i \rangle$ processes due to the collapse mechanism.

For example, if both particles have positive mass squared and the separation between particle 1 and 2 is spacelike, then from equations (\ref{A1prc}) and (\ref{A2prc}), we see that there is a drift towards a decrease in particle mass squared which is proportional to the invariant separation between particles; and from equation (\ref{A3prc}) we see that there is a drift towards a decrease in invariant separation (as though the particles are attracted to one another). If the separation between particles becomes timelike the particles' mass squared will then begin to increase on average. These effects could be negligibly small.

These considerations also raise the possibility that the particle mass squared could become negative (indicating tachyonic particle behaviour). In practice we would not expect this to occur. If there is little variance in the particle mass squared (as a result of the collapse process) then as the particle mass squared approaches zero the only way for the dispersion relation to vary continuously is for the particle energy to become infinite. This should prevent the mass squared of particles from becoming negative. Also if the collapse strengths $\lambda_1$ and $\lambda_2$ are larger than $\lambda_3$ then we expect that the particle separations will become timelike (before the mass squared become negative) which by (\ref{A1prc}) and (\ref{A2prc}) will cause the mass squared to increase.

\section{N Particles}
We now consider the case of a system of $N$ distinguishable particles. We assume that the state obeys an equation of the form (\ref{gensde}), where the set of collapse operators $\hat{\bf A}$ comprise operators of the form 
\begin{align}\label{68}
\hat{A}_{ij} =\left(\hat{x}_i - \hat{x}_j\right)^2 -  \left(\hat{t}_i - \hat{t}_j\right)^2.
\end{align}
for all particle pairs $i,j$, along with operators of the form $\hat p_i^2 - {\hat E_i}^2$ for all particles $i$. These latter operators will fix the mass states of particles. We focus our analysis on the $A_{ij}$ operators for which the density matrix $\hat \varrho$ satisfies
\begin{align}\label{69}
\frac{d}{ds}\hat{\varrho}(s)   = -i\left[ \hat{H}, \hat{\varrho}(s)\right] -\frac{\lambda}{2}\sum_{i < j} 
\left[\hat{A}_{ij},\left[\hat{A}_{ij},\varrho(s)\right]\right]
\end{align}
(we assume a common collapse strength $\lambda$).
The position-time eigenstates are eigenstates of the $A_{ij}$ operators. Let us write
\begin{align}\label{70}
|\{x_i,t_i\}\rangle  = |x_{1},t_1; x_2,t_2;\ldots ;x_N, t_N\rangle.
\end{align}
Since the set $\{ \hat{A}_{ij}\}$ all commute amongst each other, then if we ignore the Hamiltonian part of the master equation we have
\begin{align}\label{71}
\frac{d}{ds}\langle \{x_i,t_i\} |\hat{\varrho}(s)| \{x'_i,t'_i\}\rangle = -\frac{\lambda}{2}\sum_{i<j}
(A_{ij} - A'_{ij})^2 \langle \{x_i,t_i\} |\hat{\varrho}(s)|\{x'_i,t'_i\}\rangle,
\end{align}
with solution
\begin{align}\label{72}
\langle \{x_i,t_i\} |\hat{\rho}(s)| \{x'_i,t'_i\}\rangle = \exp\left\{
-\frac{\lambda}{2}(s-s_0)\sum_{i<j}
(A_{ij} - A'_{ij})^2
\right\}
\langle  \{x_i,t_i\} |\hat{\rho}(s_0)| \{x'_i,t'_i\}\rangle.
\end{align}
This implies that the state will tend towards a state with a unique value of the set $\{A_{ij}\}$. In other words, for any pair of particles $i$ and $j$, with coordinates $x_i,t_i$ and $x_j, t_j$ we expect their invariant separation 
\begin{align}\label{73}
(x_i - x_j)^2 -  (t_i -t_j)^2,
\end{align}
to tend towards a definite value.

If $N$ is sufficiently large, this will lead to a definite configuration of particles in space and time: The number of pairs of particles is 
\begin{align}\label{74}
{N\choose 2} = \frac{N!}{2!(N-2)!}.
\end{align}
For each pair, there is one constraint on the set of particle coordinates $\{x_k,t_k\}$ in the form of a definite value for the proper separation of that pair. Since the number of coordinates in this set is $2(N-1)$ (taking away one particle whose coordinates  serves as the position relative to which all other particle positions are measured) then in order to fix all these coordinates we require
\begin{align}\label{75}
\frac{N!}{2!(N-2)!} \geq 2(N-1) \implies N\geq 4.
\end{align}
This means that if the number of particles is greater than or equal to 4 then this collapse model will tend to localise the spacetime coordinates of all particles relative to one another. The particles will be localised in space and in time.

We also note that equations (\ref{econ1}) and (\ref{econ2}) describing conservation of expected energy straightforwardly generalise to the $N$-particle model and therefore energy is conserved on average.

To account for particle interactions we might include a potential term in the Hamiltonian, $V = V(|(x_i - x_j)^2 -  (t_i -t_j)^2|)$.

\section{Concluding Remarks}
The key ideas of our model are that time is treated as a quantum variable for each particle, with the CSL  Schr\"odinger equation describing evolution through space-time via an evolution parameter $s$. 

We have shown how the stochastic dynamics can lead to particles with well-defined mass, and how superpositions of different particle configurations are suppressed leading to states of definite configurations with Born rule probabilities. 

Our model describes relativistic particles and is Poincar\'e covariant. In contrast to many previous collapse models involving localisation of particle positions, in this model, the energy is conserved in stochastic expectation.

The dynamics is mapped onto spacetime by assuming that each value of $s$ has equal weight and for each $s$ there is some distribution of states of $x$ and $t$. This is done separately for each particle and so particles are assumed to be distinguishable. All information about particle interactions and entanglement is handled at the state level.


\end{document}